\documentclass[reqno,11pt]{amsart}
\usepackage{multind}
\usepackage{hyperref}
\usepackage{graphicx}
\usepackage{amscd}
\usepackage{slashed}
\usepackage[mathscr]{eucal}

\input amssym.def
\input amssym.tex

\def\goth{\frak}
\def\double{\mathbb}

\def\cc{{\double C}}

\def\rr{{\double R}}
\def\zz{{\double Z}}

\def\da{{\partial}_{\!\mbox{\tiny A}}}
\def\db{{\partial}_{\mbox{\tiny B}}}
\def\dD{{\partial}_{\mbox{\tiny D}}}

\def\DDD{{/\!\!\!\!D}}
\def\dda{{/\!\!\!\partial}_{\!\!\mbox{\tiny A}}}
\def\ddb{{/\!\!\!\partial}_{\!\mbox{\tiny B}}}
\def\ddD{{/\!\!\!\partial}_{\!\mbox{\tiny D}}}

\def\ot{\otimes}
\def\op{\oplus}

\def\bb{\begin{eqnarray}}
\def\ee{\end{eqnarray}}

\newcommand{\beq}{\begin{equation}}
\newcommand{\eeq}{\end{equation}}

\newtheorem{definition}{Definition}[section]

\newtheorem{proposition}{Proposition}[section]

\newcommand{\Proof}{\begin{proof}}
\newcommand{\QED}{\end{proof} \noindent}

\title[Dirac operators and dynamical torsion]{Dynamical torsion in view of a distinguished class of Dirac operators}

\author[J.\ Tolksdorf]{J\"urgen Tolksdorf \\ \\ \today}
\address{Max Planck Institute for Mathematics in the Sciences, Leipzig, Germany}
\email{Juergen.Tolksdorf@mis.mpg.de}
\thanks{The research leading to these results has received funding from the
European Research Council under the European Union's Seventh Framework
Programme (FP7/2007-2013) / ERC grant agreement n$^\circ$~267087.}

\begin{document}
\maketitle

\begin{abstract}
In this paper we discuss geometric torsion in terms of a distinguished class of Dirac operators. 
We demonstrate that from this class of Dirac operators a variational problem for torsion can be derived similar to that of Yang-Mills gauge theory. As a consequence, one ends up with propagating torsion
even in vacuum as opposed to Einstein-Cartan theory.
\end{abstract}

\tableofcontents




\section{Introduction}
A connection on the frame bundle of any smooth manifold $M$ is known to yield the two independent geometrical concepts of {\it curvature} and {\it torsion}. There are various (but equivalent) approaches to the torsion of a connection, depending on the geometrical setup. For instance, the torsion of a connection 
$\nabla\equiv\nabla^{\mbox{\tiny TM}}$ on the tangent bundle of $M$ may be defined by
\beq
\label{tor}
\tau_{\mbox{\tiny$\nabla$}} := d_ {\mbox{\tiny$\nabla$}}\goth{Id}\in\Omega^2(M,TM)\,.
\eeq 
Here, $d_ {\mbox{\tiny$\nabla$}} $ denotes the exterior covariant derivative with respect to 
$\nabla^ {\mbox{\tiny TM}}$. The canonical one-form $\goth{Id}\in\Omega^1(M,TM)$ is defined as 
$\goth{Id}_x(v) := v$ for all tangent vectors $v\in T_x M$ and $x\in M$. This canonical one-form corresponds to the soldering form on the frame bundle of $M$.

In general relativity a field equation for the torsion is obtained by the so-called ``Palatini formalism'',
where the metric and the connection on the tangent bundle are regarded as being independent from 
each other (c.f. \cite{Misner et al}). The resulting field equation for torsion in Einstein's theory of gravity is known to be given by 
\beq
\label{torpalatini}
\tau_{\mbox{\tiny$\nabla$}} = \lambda_{\mbox{\tiny grav}}\,{\goth j}_{\mbox{\tiny spin}}\,,
\eeq
where the real coupling constant $\lambda_{\mbox{\tiny grav}}$ is proportional to the gravitational 
constant. The so-called ``spin-current'' ${\goth j}_{\mbox{\tiny spin}}\in\Omega^2(M,TM)$ is obtained 
by the variation of the action functional that dynamically describes matter fields with respect to the metric connection. Usually, ordinary bosonic matter does not depend on the metric connection as, for instance, described by the Standard Model. According to \eqref{torpalatini} the connection is thus provided by the Levi-Civita connection. This holds true, in particular, when matter is disregarded. When matter is defined in terms of spinor fields, as in the case of the Dirac action, then the right-hand side of \eqref{torpalatini} may be non-vanishing. This is usually rephrased by the statement 
\begin{center}
``{\it Spin is the source of torsion}''. 
\end{center}
However, even in this case torsion is not propagating in space-time since \eqref{torpalatini} is purely algebraic relation between torsion and matter. Furthermore, a torsion-free connection provides a sufficient condition for the ordinary Dirac action to be real. More precisely, with respect to the identification
$\Omega^2(M,TM)\simeq\Omega^1(M,{\rm End}(TM))$ a necessary and sufficient condition for the 
realness of the ordinary Dirac action is given by (c.f. \cite{Goeckeler et al})
\beq
tr\tau_{\mbox{\tiny$\nabla$}}(v) = 0\quad\big(v\in TM\big)\,.
\eeq 

In this work, we discuss torsion within the framework of general relativity which is different from the
ordinary Palatini formalism.  We obtain field equations for the torsion from a functional that looks similar 
to the Einstein-Hilbert-Yang-Mills-Dirac action. This functional is derived from a certain class of Dirac operators. The geometrical background of these Dirac operators is basically dictated by the reality condition imposed on the action including (full) torsion. Furthermore, this class of Dirac operators fits 
well with those giving rise to Einstein's theory of gravity, ordinary Yang-Mills-Dirac theory and non-linear 
$\sigma-$models, as discussed in \cite{Tolksdorf1}. 

From a physics point of view torsion provides an additional degree of freedom to Einstein's theory of
gravity. In the latter the action of a gravitational field is described in terms of the curvature of a smooth 
four dimensional manifold $M$. Even more, this curvature is assumed to be uniquely determined by the Levi-Civita connection on the tangent bundle of $M$ with fiber metric $g_{\mbox{\tiny M}}$. In other
words, the geometrical model behind Einstein's theory of gravity is known to be given by a 
smooth (orientable) Lorentzian four-manifold $(M,g_{\mbox{\tiny M}})$ of signature $s=\pm 2$ (resp. a
diffeomorphism class thereof). For a ``space-time'' $(M,g_{\mbox{\tiny M}})$ to be physically admissible
the metric field $g_{\mbox{\tiny M}}$ has to fulfill the Einstein equation of gravity (and maybe topological 
restriction on $M$, like global hyperbolicity), whereby the source of gravity is given by the energy-momentum current of matter. The latter is either phenomenologically described by a mass density or 
in terms of matter fields (i.e. sections of certain vector bundles over space-time). This holds true, 
especially, if the spin of matter is taken into account. In this case, matter is geometrically described by
(Dirac) spinor fields. According to the above mentioned statement about the relation between spin and
torsion a huge variety of generalizations of Einstein's theory of gravity including torsion has been proposed over the last decades, going under the name ``Einstein-Cartan theory'', ``Poincare gauge gravity'', ``teleparallel gravity'', etc. (see, for instance, \cite{Hehl et al}, as well as the more recent essay 
\cite{Hehl2 et al} and the references cited therein). For higher order gravity with (propagating) torsion discussed within the realm of Connes' non-commutative geometry we refer to \cite{Hanisch et al}, \cite{Pfaeffle et al} and the references cited therein. In these references only the totally anti-symmetric 
part of the torsion tensor is proposed to propagate according to a Procca like field equation. In contrast, 
in this work we discuss Yang-Mills like equations for the complete torsion tensor within Einstein's theory of 
gravity. Moreover, in the approach discussed here the (so-called ``strong'') equivalence principle still holds because torsion is assumed to couple only to the ``internal degrees of freedom'' of the fermions (see below).

In some of the above mentioned approaches torsion does not propagate, whereas other approaches propose only torsion but no curvature. In any case, spin is considered to be the source of torsion. Since spin is fundamental torsion plays also a prominent role in (super-)string theory (see, for instance, \cite{Bars et al} and \cite{Li et al}). For an overview about the role of torsion in theoretical physics we refer to \cite{Gogala}. Also, we refer to \cite{Torsion in Physical Theories} as a reasonable source of references to the issue. From a physics perspective it is speculated that torsion might contribute to dark energy, whose existence seems experimentally confirmed by the observed acceleration of the universe.

The approach to a dynamical torsion presented in this work is different, for it starts out with Rarita-Schwinger fermions to which torsion minimally couples. These fermions are geometrically modeled 
by sections of a twisted spinor bundle, where the ``inner degrees of freedom'' are generated by the
(co-)tangent bundle of the underlying manifold. As a consequence, the resulting coupling to torsion completely parallels that of spinor-electrodynamics. Since the known matter is geometrically described by Dirac spinors, the Rarita-Schwinger fermions may serve as a geometrical model to physically describe 
dark matter (or parts thereof). Accordingly, the energy momentum of torsion (the underlying gauge field) may contribute to dark energy. The coupling constant between the Rarita-Schwinger and the torsion field, however, is a free parameter like in ordinary gauge theory, although the underlying action is derived from ``first (geometrical) principles''. Hence, this gauge coupling constant is an additional free parameter as opposed to ordinary Einstein-Cartan theory.

This paper is organized as follows: We start out with a summary of the necessary geometrical 
background of Dirac operators in terms of general Clifford module bundles. Afterwards, we discuss 
torsion in the context of a distinguished class of Dirac operators which give rise to field equations similar 
to Einstein-Dirac-Yang-Mills equations.

\section{Geometrical background}
The geometrical setup presented fits well with that already discussed in
\cite{Tolksdorf1} for non-linear $\sigma-$models and Yang-Mills theory.
For the convenience of the reader we briefly summarize the basic geometrical background. 
In particular, we present the basic features of Dirac operators of simple type. This class of Dirac 
operators will play a fundamental role in our discussion. For details we refer to \cite{Tolksdorf2} (or
\cite{Tolksdorf1}) and \cite{Atiyah et al}, as well as to \cite{Berline et al} which serves as a kind of 
``standard reference'' for what follows.

In the sequel, $(M,g_{\mbox{\tiny M}})$ always denotes a smooth orientable (semi-)Riemannian
manifold of finite dimension $n\equiv p + q$. The index of the (semi-)Riemannian metric 
$g_{\mbox{\tiny M}}$ is $s\equiv p - q\not\equiv 1\text{~mod~}4$. The {\it bundle of exterior forms} of degree $k\geq 0$ is denoted by $\Lambda^{\!k}T^\ast\!M\rightarrow M$ with its canonical projection. Accordingly, the {\it Grassmann bundle} is given by 
$\Lambda T^\ast\!M \equiv \bigoplus_{k\geq 0}\Lambda^{\!k}T^\ast\!M\rightarrow M$. It naturally
inherits a metric denoted by $g_{\mbox{\tiny$\Lambda{\rm M}$}}$, such that the direct sum is orthogonal 
and the restriction of $g_{\mbox{\tiny$\Lambda{\rm M}$}}$ to degree one equals to the fiber metric 
$g^\ast_{\mbox{\tiny M}}$ of the cotangent bundle $T^\ast\!M\rightarrow M$.

The mutually inverse {\it musical isomorphisms} in terms of $g_{\mbox{\tiny M}}$ 
(resp. $g^\ast_{\mbox{\tiny M}}$) are denoted by $\phantom{x}^{\flat/\sharp}:\,TM\simeq T^\ast\!M$, such that, for instance, $g_{\mbox{\tiny M}}(u,v) = g^\ast_{\mbox{\tiny M}}(u^\flat,v^\flat)$ for all 
$u,v\in TM$.

The {\it Clifford bundle} of $(M,g_{\mbox{\tiny M}})$ is denoted by 
$\pi_{\mbox{\tiny Cl}}:\,Cl_{\mbox{\tiny M}} = Cl^+_{\mbox{\tiny M}}\op Cl^-_{\mbox{\tiny M}}\rightarrow M$.
Its canonical ``even/odd'' grading involution is $\tau_{\mbox{\tiny Cl}}\in{\rm End}(Cl_{\mbox{\tiny M}})$.
The hermitian structure $\big<\cdot,\cdot\big>_{\!\mbox{\tiny Cl}}$ is the one induced by the metric 
$g_{\mbox{\tiny M}}$ due to the canonical linear isomorphism between the Clifford and the Grassmann bundle (c.f. \eqref{symbolmap}, below). We always assume the Clifford and the Grassmann bundle to be generated by the cotangent bundle of $M$. Notice that in the sequel the notion of ``hermitian structure'' does not necessarily imply a positive definite fiber metric. In fact, the signature of the fiber metric may depend on the signature of the underlying metric $g_{\mbox{\tiny M}}$, like in the case of the Clifford bundle associated to $(M,g_{\mbox{\tiny M}})$.

Throughout the present work we always identify on $(M,g_{\mbox{\tiny M}})$ the vector bundle
$\pi_{\mbox{\tiny$\Lambda{\rm T}^\ast\!{\rm M}$}}|_{\mbox{\tiny$\Lambda^{\!2}{\rm T}^\ast\!{\rm M}$}}:\,\Lambda^{\!2}T^\ast\!M\rightarrow M$ with the (Lie algebra) bundle 
$\pi_{\mbox{\tiny so}}:\,so(TM)\rightarrow M$ of the $g_{\mbox{\tiny M}}$ skew-symmetric endomorphisms on the tangent bundle $\pi_{\mbox{\tiny TM}}:\,TM\rightarrow M$ due to the canonical linear (bundle) isomorphism (over the identity on $M$)
\beq
\begin{split}
\Lambda^{\!2}T^\ast\!M &\stackrel{\simeq}{\longrightarrow} so(TM)\subset{\rm End}(TM)\\
\omega &\mapsto \Omega\,,
\end{split}
\eeq 
where for all $u,v\in TM$: $g_{\mbox{\tiny M}}(u,\Omega(v)) := \omega(u,v)$. Accordingly, we always take advantage of the induced isomorphism
\beq
\begin{split}
\Lambda^{\!2}T^\ast\!M\ot TM &\stackrel{\simeq}{\longrightarrow} T^\ast\!M\ot so(TM)\\
\omega\ot v &\mapsto v^\flat\ot\Omega\,.
\end{split}
\eeq 

A smooth complex vector bundle $\pi_{\mbox{\tiny$\mathcal{E}$}}:\,\mathcal{E}\rightarrow M$ of
rank $rk(\mathcal{E})\geq 1$ is called a {\it Clifford module bundle}, provided there is a {\it Clifford map}. That is, there is a smooth linear (bundle) map (over the identity on $M$)
\beq
\label{cliffmap}
\begin{split}
\gamma_{\mbox{\tiny$\mathcal{E}$}}:\,T^\ast\!M &\longrightarrow {\rm End}(\mathcal{E})\\
\alpha &\mapsto \gamma_{\mbox{\tiny$\mathcal{E}$}}(\alpha)\,,
\end{split}
\eeq
satisfying $\gamma_{\mbox{\tiny$\mathcal{E}$}}(\alpha)^2 = 
\epsilon g^\ast_{\mbox{\tiny M}}(\alpha,\alpha){\rm Id}_{\mbox{\tiny$\mathcal{E}$}}$. Here,
$\epsilon\in\{\pm 1\}$ depends on how the {\it Clifford product} is defined. That is,
$\alpha^2 := \pm g^\ast_{\mbox{\tiny M}}(\alpha,\alpha)1_{\mbox{\tiny Cl}}\in Cl_{\mbox{\tiny M}}$,
for all $\alpha\in T^\ast\!M\subset Cl_{\mbox{\tiny M}}$ and $1_{\mbox{\tiny Cl}}\in Cl_{\mbox{\tiny M}}$ 
denotes the unit element. 

To emphasize the module structure we write
\beq
\label{cliffmodbdl}
\pi_{\mbox{\tiny$\mathcal{E}$}}:\,
(\mathcal{E},\gamma_{\mbox{\tiny$\mathcal{E}$}})\longrightarrow(M,g_{\mbox{\tiny M}})\,.
\eeq

The bundle \eqref{cliffmodbdl} is called an {\it odd hermitian Clifford module bundle},
provided it is $\zz_2-$graded, with grading involution $\tau_{\mbox{\tiny$\mathcal{E}$}}$, and endowed
with an hermitian structure $\big<\cdot,\cdot\big>_{\!\mbox{\tiny$\mathcal{E}$}}$, such that
$\gamma_{\mbox{\tiny$\mathcal{E}$}}\circ\tau{\mbox{\tiny$\mathcal{E}$}} =
-\tau_{\mbox{\tiny$\mathcal{E}$}}\circ\gamma_{\mbox{\tiny$\mathcal{E}$}}$ and both the grading 
involution and Clifford action are either hermitian or skew-hermitian. In what follows, \eqref{cliffmodbdl}
always means an odd hermitian Clifford module. Notice that by ``hermitian'' we simply mean a fiber 
metric of arbitrary signature. The archetype example of an odd hermitian Clifford module over
$(M,g_{\mbox{\tiny M}})$ is provided by the corresponding Clifford bundle.
 
The linear map
\beq
\label{quantmap}
\begin{split}
\delta_\gamma:\,\Omega(M,{\rm End}(\mathcal{E})) &\longrightarrow 
\Omega^0(M,{\rm End}(\mathcal{E}))\\
\omega\equiv\alpha\otimes\goth{B} &\mapsto {/\!\!\!\!\omega}\equiv
\gamma_{\mbox{\tiny$\mathcal{E}$}}\big(\sigma_{\!\mbox{\tiny Ch}}^{-1}(\alpha)\big)\circ\goth{B}
\end{split}
\eeq
is called the {\it ``quantization map''}. It is determined by the linear isomorphism called {\it symbol map}:
\beq
\label{symbolmap}
\begin{split}
\sigma_{\!\mbox{\tiny Ch}}:\,Cl_{\mbox{\tiny M}} &\stackrel{\simeq}{\longrightarrow} \Lambda T^\ast\!M\\
\goth{a} &\mapsto \Gamma_{\!\mbox{\tiny Ch}}(\goth{a})1_{\mbox{\tiny$\Lambda$}}\,.
\end{split}
\eeq
Here, $1_{\mbox{\tiny$\Lambda$}}\in\Lambda T^\ast\!M$ is the unit element. The homomorphism 
$\Gamma_{\!\mbox{\tiny Ch}}:\,Cl_{\mbox{\tiny M}}\rightarrow{\rm End}(\Lambda T^\ast M)$ 
is given by the canonical Clifford map:
\beq
\begin{split}
\gamma_{\mbox{\tiny Cl}}:\,T^\ast\!M &\longrightarrow {\rm End}(\Lambda T^\ast\!M)\\
v &\mapsto \left\{
  \begin{array}{ccc}
    \Lambda T^\ast\!M & \longrightarrow & \hspace{-2cm}\Lambda T^\ast\!M \\ 
    \omega & \mapsto & \epsilon int(v)\omega + ext(v^\flat)\omega\,,
  \end{array}
\right.
\end{split}
\eeq
where, respectively, $``int"$ and $``ext"$ indicate ``interior'' and ``exterior'' multiplication.

When restricted to $\Omega^1(M,{\rm End}(\mathcal{E}))$ the quantization map \eqref{quantmap}
has a canonical right-inverse given by
\beq
\begin{split}
ext_{\mbox{\tiny$\Theta$}}:\,\Omega^0(M,{\rm End}(\mathcal{E})) &\longrightarrow
\Omega^1(M,{\rm End}(\mathcal{E}))\\
\Phi &\mapsto \Theta\Phi\,,
\end{split}
\eeq
where the {\it canonical one-form} $\Theta\in\Omega^1(M,{\rm End}(\mathcal{E}))$ is given by
$\Theta(v) := \frac{\varepsilon}{n}\gamma_{\mbox{\tiny$\mathcal{E}$}}(v^\flat)$, for all $v\in TM$.
The associated projection operators are 
${\goth p} \equiv ext_{\mbox{\tiny$\Theta$}}\circ\delta_\gamma|_{\Omega^1}$ and 
${\goth q} := {\rm Id}_{\Omega^1} - {\goth p}$, such that
\beq
\label{oneformdecomp}
\Omega^1(M,{\rm End}(\mathcal{E})) =
{\goth p}\big(\Omega^1(M,{\rm End}(\mathcal{E}))\big)\op
{\goth q}\big(\Omega^1(M,{\rm End}(\mathcal{E}))\big)\,.
\eeq

Notice that for any connection on a Clifford module bundle the first order operator 
$\mathcal{T}_{\!\mbox{\tiny$\nabla$}}\equiv{\goth q}(\nabla^{\mbox{\tiny$\mathcal{E}$}}):\,
\Omega^0(M,\mathcal{E})\rightarrow\Omega^1(M,\mathcal{E}),\,\psi\mapsto 
\nabla^{\mbox{\tiny$\mathcal{E}$}}\psi - \Theta\big(\,{/\!\!\!\!\nabla^{\mbox{\tiny$\mathcal{E}$}}}\!\psi\big)$ 
is the associated {\it twister operator}. Here, $\,{/\!\!\!\!\nabla^{\mbox{\tiny$\mathcal{E}$}}}\equiv\delta_\gamma(\nabla^{\mbox{\tiny$\mathcal{E}$}})$ denotes the Dirac operator associated to the connection 
(see below).

A (linear) connection on a Clifford module bundle is called a {\it Clifford connection} if the corresponding covariant derivative $\nabla^{\mbox{\tiny$\mathcal{E}$}}$ ``commutes'' with the Clifford map 
$\gamma_{\mbox{\tiny$\mathcal{E}$}}$ in the following sense:
\beq
\label{cliffcon1}
[\nabla^{\mbox{\tiny$\mathcal{E}$}}_{\!\!X},\gamma_{\mbox{\tiny$\mathcal{E}$}}(\alpha)] = 
\gamma_{\mbox{\tiny$\mathcal{E}$}}\big(\nabla^{\mbox{\tiny$T^\ast\!M$}}_{\!\!X}\alpha\big)\qquad
\big(X\in\goth{S}ec(M,TM),\;\alpha\in\goth{S}ec(M,T^\ast\!M)\big)\,.
\eeq
Here, $\nabla^{\mbox{\tiny$T^\ast\!M$}}$ is the Levi-Civita connection on the co-tangent bundle
with respect to $g_{\mbox{\tiny M}}^\ast$.

Equivalently, a connection on a Clifford module bundle is a Clifford connection if and only if it fulfills:
\beq
\nabla_{\!\!X}^{\mbox{\tiny$T^\ast\!M\!\ot\!{\rm End}(\mathcal{E})$}}\Theta = 0\qquad
\big(X\in\goth{S}ec(M,TM)\big)\,.
\eeq
Apparently, Clifford connections provide a distinguished class of connections on any Clifford module bundle.

We denote Clifford connections by $\da$. This notation is used because Clifford connections are  
parametrized by a family of locally defined one-forms $A\in\Omega^1(U,{\rm End}_\gamma(\mathcal{E}))$.
Here, $U\subset M$ will always denote a connected open subset and 
${\rm End}_\gamma(\mathcal{E})\subset{\rm End}(\mathcal{E})$ is the total space of the 
algebra bundle of endomorphisms which commute with the Clifford action provided by  
$\gamma_{\mbox{\tiny$\mathcal{E}$}}$. 

We call in mind that a {\it Dirac operator} $\DDD$ on a Clifford module bundle is a first order differential operator acting on sections $\psi\in\goth{S}ec(M,\mathcal{E})$, such that
$[\,\DDD,df]\psi = \gamma_{\mbox{\tiny$\mathcal{E}$}}(df)\psi$ for all smooth functions 
$f\in\mathcal{C}^\infty(M)$.
The set of all Dirac operators on a given Clifford module bundle is denoted by 
$\goth{Dir}(\mathcal{E},\gamma_{\mbox{\tiny$\mathcal{E}$}})$. It is an affine set over the vector space 
$\Omega^0(M,{\rm End}(\mathcal{E}))$. Moreover, Dirac operators are defined to be {\it odd} operators 
on odd (hermitian) Clifford module bundles: 
$\,\DDD\tau_{\mbox{\tiny$\mathcal{E}$}} = -\tau_{\mbox{\tiny$\mathcal{E}$}}\,\DDD$ . In this case, the underlying vector space reduces to $\Omega^0(M,{\rm End}^-(\mathcal{E}))$.

We call the Dirac operator ${/\!\!\!\!\nabla}^{\mbox{\tiny$\mathcal{E}$}}\equiv
\delta_\gamma(\nabla^{\mbox{\tiny$\mathcal{E}$}})$ the {\it ``quantization''} of a connection 
$\nabla^{\mbox{\tiny$\mathcal{E}$}}$ on a Clifford module bundle \eqref{cliffmodbdl}. Let
$e_1,\ldots,e_n\in\goth{S}ec(U,TM)$ be a local frame and $e^1,\ldots,e^n\in\goth{S}ec(U,T^\ast\!M)$
its dual frame. For $\psi\in\goth{S}ec(M,\mathcal{E})$ one has
\beq
{/\!\!\!\!\nabla}^{\mbox{\tiny$\mathcal{E}$}}\!\psi := 
\sum_{k = 1}^n\delta_\gamma(e^k)\nabla_{\!\!e_k}^{\mbox{\tiny$\mathcal{E}$}}\psi =
\sum_{k = 1}^n\gamma_{\mbox{\tiny$\mathcal{E}$}}(e^k)\nabla_{\!\!e_k}^{\mbox{\tiny$\mathcal{E}$}}\psi\,,
\eeq
where the canonical embedding $\Omega(M)\hookrightarrow\Omega(M,{\rm End}(\mathcal{E})),\;\omega\mapsto \omega\equiv\omega\ot{\rm Id}_{\mbox{\tiny$\mathcal{E}$}}$ is taken into account.

Every Dirac operator has a canonical first-order decomposition:
\beq
\label{1dirdecomp}
\DDD = \ddb + \Phi_{\mbox{\tiny D}}\,.
\eeq
Here, $\db$ denotes the (covariant derivative of the) {\it Bochner connection} that is defined by $\,\DDD$ 
as
\beq
\label{bochcon}
2ev_g\big(df,\db\psi\big) := \epsilon\big([\,\DDD^2,f] - \delta_{\!g} df\big)\psi\qquad
\big(\psi\in\goth{S}ec(M,\mathcal{E})\big)\,,
\eeq
with $ev_g$'' being the evaluation map with respect to $g_{\mbox{\tiny M}}$ and
$\delta_{\!g}$ the dual of the exterior derivative (see \cite{Berline et al}).
 
The zero-order section $\Phi_{\mbox{\tiny D}} := \,\DDD - \ddb\in\goth{S}ec(M,{\rm End}(\mathcal{E}))$ is thus also 
uniquely determined by $\,\DDD$. We call the Dirac operator $\ddb$ the {\it``quantized Bochner connection''}. 

Since the set $\goth{Dir}(\mathcal{E},\gamma_{\mbox{\tiny$\mathcal{E}$}})$ is an affine space, 
every Dirac operator can be written as
\beq
\label{noncandecomp}
\DDD = \dda + \Phi\,.
\eeq
However, this decomposition is far from being unique. The
section $\Phi\in\goth{S}ec(M,{\rm End}(\mathcal{E}))$ depends on the chosen Clifford connection $\da$. 
In general, a Dirac operator does not uniquely determine a Clifford connection.

\begin{definition}
A Dirac operator $\,\DDD\in{\goth Dir}(\mathcal{E},\gamma_{\mbox{\tiny$\mathcal{E}$}})$ is said to be of ``simple type'' provided that $\Phi_{\!\mbox{\tiny D}} = 
\,\DDD - \ddb\in{\goth S}ec(M,{\rm End}(\mathcal{E}))$ anti-commutes with the Clifford action:
\beq
\Phi_{\!\mbox{\tiny D}}\gamma_{\mbox{\tiny$\mathcal{E}$}}(\alpha) = 
-\gamma_{\mbox{\tiny$\mathcal{E}$}}(\alpha)\Phi_{\!\mbox{\tiny D}}\qquad\big(\alpha\in T^\ast\!M\big)\,.
\eeq
\end{definition}

It follows that a Dirac operator of simple type uniquely determines a Clifford connection $\da$ together with a zero-order operator $\phi_{\mbox{\tiny D}}\in\goth{S}ec(M,{\rm End}_\gamma(\mathcal{E}))$, such that
(c.f. \cite{Tolksdorf2})
\beq
\label{simpletypedop}
\DDD = \dda + \tau_{\mbox{\tiny$\mathcal{E}$}}\phi_{\mbox{\tiny D}}\,.
\eeq
These Dirac operators play a basic role in the geometrical description of the Standard Model
(c.f. \cite{Tolksdorf2}). They are also used in the context of the family index theorem (see, for instance, 
\cite{Berline et al}).  Apparently, Dirac operators of simple type provide a natural generalization of 
quantized Clifford connections. Indeed, they build the biggest class of Dirac operators such that the corresponding Bochner connections are also Clifford connections. Notice that 
$\phi_{\mbox{\tiny D}}\in\goth{S}ec(M,{\rm End}^-_\gamma(\mathcal{E}))$ for odd (hermitian) Clifford modules. 

Every Dirac operator is known to have a unique {\it second order decomposition}
\beq
\label{2nddirdecomp}
\DDD^2 = \triangle_{\mbox{\tiny B}} + V_{\mbox{\tiny D}}\,,
\eeq
where the {\it Bochner-Laplacian} (or ``trace Laplacian'') is given in terms of the Bochner connection as
$\triangle_{\mbox{\tiny B}} := \epsilon ev_g\big(\db^{\mbox{\tiny$T^\ast\!M\!\ot\!\mathcal{E}$}}\circ\db\big)$
(c.f. \cite{Berline et al}). 

The trace of the zero-order operator $V_{\mbox{\tiny D}}\in\goth{S}ec(M,{\rm End}(\mathcal{E}))$ explicitly reads (c.f. \cite{Tolksdorf2}): 
\beq
\label{trdirpot}
tr_{\!\mbox{\tiny$\mathcal{E}$}}V_{\!\mbox{\tiny D}} = 
tr_{\!\gamma}\!\left(curv(\,\DDD) - \epsilon ev_{g}(\omega_{\mbox{\tiny D}}^2)\right) - 
\varepsilon\delta_{\!g}\big(tr_{\!\mbox{\tiny$\mathcal{E}$}}\omega_{\mbox{\tiny D}}\big)\,,
\eeq
where $curv(\,\DDD)\in\Omega^2(M,{\rm End}(\mathcal{E}))$ denotes the curvature of the Dirac 
connection of $\,\DDD\in\goth{Dir}(\mathcal{E},\gamma_{\mbox{\tiny$\mathcal{E}$}})$ and
$tr_{\!\gamma} := tr_{\!\mbox{\tiny$\mathcal{E}$}}\circ\delta_{\gamma}$ the ``quantized trace''.
The {\it Dirac connection} of $\,\DDD$ is defined by $\dD := \db + \omega_{\mbox{\tiny D}}$, with
the {\it ``Dirac form''} $\omega_{\mbox{\tiny D}}\equiv ext_{\mbox{\tiny$\Theta$}}\Phi_{\mbox{\tiny D}}\in
\Omega^1(M,{\rm End}^+(\mathcal{E}))$. The Dirac connection has the property that it is uniquely determined by $\,\DDD$, as opposed to the decomposition \eqref{noncandecomp} in terms of quantized Clifford connections. Furthermore, one has $\ddD = \,\DDD$. Notice that in general $\ddb\not=\,\DDD$.
Indeed, $\ddb = \,\DDD$ holds true if and only if $\,\DDD$ is a quantized Clifford connection. Only in this case, the three notions of Dirac-, Bochner- and Clifford connection coincide.

Let $M$ be closed compact. We call the canonical mapping
\beq
\label{diract}
\begin{split}
\mathcal{I}_{\mbox{\tiny D}}:\, \goth{Dir}(\mathcal{E},\gamma_{\mbox{\tiny$\mathcal{E}$}}) &\rightarrow\cc\\
\DDD &\mapsto \int_M \ast tr_{\!\mbox{\tiny$\mathcal{E}$}}V_{\!\mbox{\tiny D}}
\end{split}
\eeq
the {\it ``universal Dirac action''} and
\beq
\label{totdiract}
\begin{split}
\mathcal{I}_{\mbox{\tiny D,tot}}:\, \goth{Dir}(\mathcal{E},\gamma_{\mbox{\tiny$\mathcal{E}$}})\times
\goth{S}ec(M,\mathcal{E}) &\rightarrow\cc\\
(\,\DDD,\psi) &\mapsto \int_M \ast\big(\big<\psi,\,\DDD\psi\big>_{\!\mbox{\tiny$\mathcal{E}$}} +
tr_{\!\mbox{\tiny$\mathcal{E}$}}V_{\!\mbox{\tiny D}}\big)
\end{split}
\eeq
the {\it ``total Dirac action''}. Here, ``$\ast$'' is the Hodge map with respect to $g_{\mbox{\tiny M}}$
and a chosen orientation of $M$.

When \eqref{trdirpot} is taken into account, the universal Dirac action
\beq
\label{diract2}
\mathcal{I}_{\mbox{\tiny D}}(\,\DDD\,) =
\int_M\ast tr_{\!\gamma}\!\left(curv(\,\DDD) - 
\epsilon\,ev_{g}(\omega_{\mbox{\tiny D}}^2)\right)
\eeq
is seen to provide a natural generalization of the Einstein-Hilbert functional with a cosmological constant:
\beq
\label{ehc}
\begin{split}
\mathcal{I}_{\mbox{\tiny EHC}}(g_{\mbox{\tiny M}}) &:=
\int_M\ast\!\left(scal(g_{\mbox{\tiny M}}) + \Lambda\right)\\
&\thicksim
\int_M\ast tr_{\!\gamma}\!\left(curv({\,/\!\!\!\!\nabla^{\mbox{\tiny$\mathcal{E}$}}}) - 
\epsilon\, ev_{g}(\omega^2)\right)\,.
\end{split}
\eeq
Here, $\nabla^{\mbox{\tiny$\mathcal{E}$}}$ is a Clifford connection on a Clifford module bundle
\eqref{cliffmodbdl}. For instance, one may take the Clifford bundle and the lifted Levi-Civita connection. Furthermore, we set 
$\omega\equiv \sqrt{\frac{\epsilon n}{4}\Lambda}\,\Theta\in\Omega^1(M,{\rm End}^-(\mathcal{E}))$. 
Notice, however, that this form is odd on odd hermitian Clifford module bundles. In contrast, the Dirac form 
$\omega_{\mbox{\tiny D}}\in\Omega^1(M,{\rm End}^+(\mathcal{E}))$, which appears in \eqref{diract2}, is always even on odd hermitian Clifford module bundles. The smooth function $scal(g_{\mbox{\tiny M}})$ is the {\it scalar curvature} of the Levi-Civita connection of $g_{\mbox{\tiny M}}$ and $\Lambda\in\rr$ is the ``cosmological constant''.

Indeed, when $\,\DDD$ is a quantized Clifford connection, then the universal Dirac action \eqref{diract2} 
reduces to the Einstein-Hilbert functional. This follows from the Schr\"odinger-Lichnerowicz decomposition of the square of a quantized Clifford connection (c.f. \cite{Schroedinger}, \cite{Lichnerowicz}. In contrast, 
for Dirac operators of simple type the universal Dirac action becomes 
\beq
\label{stypediract}
\mathcal{I}_{\mbox{\tiny D}}\big(\dda + \tau_{\mbox{\tiny$\mathcal{E}$}}\phi_{\mbox{\tiny D}}\big) = 
\int_M\!\ast\big(-\mbox{\small$\epsilon\frac{rk(\mathcal{E})}{4}$}scal(g_{\mbox{\tiny M}}) + 
tr_{\!\mbox{\tiny$\mathcal{E}$}}\phi_{\mbox{\tiny D}}^2\big)\,.
\eeq
This explicit formula is a direct consequence of Lemma 4.1 and the Corollary 4.1 of 
Ref. \cite{Tolksdorf2} (see also Sec. 6 in loc. site), which generalizes the Schr\"odinger-Lichnerowicz formula to arbitrary Dirac operators. Therefore, restriction of the universal Dirac action \eqref{diract} to Dirac operators of simple type \eqref{simpletypedop} corresponds to \eqref{ehc}, where (up to numerical factors)
\beq
\Lambda = tr_{\!\mbox{\tiny$\mathcal{E}$}}\phi_{\mbox{\tiny D}}^2\,.
\eeq
Hence, for $\phi_{\mbox{\tiny D}}^\dagger = \pm\phi_{\mbox{\tiny D}}$ the cosmological constant is
given basically by the length of the section 
$\phi_{\mbox{\tiny D}}\in{\goth S}ec(M,{\rm End}_\gamma^-(\mathcal{E}))$ that is associated with the simple type Dirac operator $\,\DDD = \pm\,\DDD^\dagger\in\goth{Dir}(M,\mathcal{E})$. Indeed, whenever
the section $\phi_{\mbox{\tiny D}}$ does not depend on the metric, the variation of \eqref{stypediract}
with respect to $g_{\mbox{\tiny M}}$ will yield that $\phi_{\mbox{\tiny D}}$ has to be of constant length.
Yet, this does not imply that the section $\phi_{\mbox{\tiny D}}$ itself has to be constant. Merely this 
means that $\phi_{\mbox{\tiny D}}$ has to be a section of the sphere sub-bundle of radius $\Lambda$ 
of the hermitian vector bundle ${\rm End}_\gamma(\mathcal{E})\rightarrow M$. This is similar to what is encountered in the Higgs sector of the Standard Model, actually. When an underlying gauge group is supposed to act transitively on this sphere bundle, then the gauge symmetry becomes ``spontaneously broken'' like in the Standard Model but without the need for a Higgs potential.  

We stress that the universal (total) Dirac action is fully determined by the Dirac operator, 
like the Einstein-Hilbert functional is fully determined by the metric. In contrast, for example, the
Yang-Mills action is not a canonical functional on the set of connections for it also depends on the
chosen metric. Also, we call in mind that nowadays the functional \eqref{ehc} is considered to be more fundamental (at least on the cosmological scale) than the original Einstein-Hilbert action. In fact, the
functional \eqref{ehc} is mathematically well-motivated due to Lovelock's Theorem (c.f. \cite{Lovelock}).
Similarly, Dirac operators of simple type seem to be more profound than quantized Clifford connections.

\section{Dirac operators with torsion}
To this end let $(M,g_{\mbox{\tiny M}})$ be a (semi-)Riemannian {\it spin-}manifold of even dimension 
$n = p+q$ and signature $p-q\not\equiv 1~\text{mod}~4$. Let 
$\pi_{\mbox{\tiny S}}:\,S = S^+\op S^-\rightarrow M$ be a (complexified) spinor bundle with grading involution $\tau_{\mbox{\tiny S}}\in{\rm End}(S)$. The hermitian structure is denoted by 
$\big<\cdot,\cdot\big>_{\!\mbox{\tiny S}}$. The Clifford action is provided by the canonical Clifford map 
$\gamma_{\mbox{\tiny S}}:\,T^\ast\!M^{\mbox{\tiny$\cc$}}\rightarrow{\rm End}_{\mbox{\tiny$\cc$}}(S)\simeq Cl_{\mbox{\tiny M}}^{\mbox{\tiny$\cc$}}$. The induced Clifford action is supposed to be anti-hermitian. The Clifford action also anti-commutes with the grading involution. The grading involution is assumed to be either hermitian or anti-hermitian. Let us call in mind that by abuse of notation ``hermitian'' does not  necessarily mean in what follows positive definiteness of the fiber metric\footnote{The author would like
to thank Ch. B\"ar (University of Potsdam/Germany) for a corresponding remark.}.

We consider the twisted spinor bundle
\beq
\label{twistedspinbdl}
\pi_{\mbox{\tiny$\mathcal{E}_1$}}:\,\mathcal{E}_1:= S\ot_{\mbox{\tiny M}}TM\longrightarrow M
\eeq
with the grading involution $\tau_{\mbox{\tiny$\mathcal{E}_1$}} := 
\tau_{\mbox{\tiny S}}\!\ot{\rm Id}_{\mbox{\tiny TM}}$ and Clifford action
$\gamma_{\mbox{\tiny$\mathcal{E}_1$}} := \gamma_{\mbox{\tiny S}}\!\ot\!{\rm Id}_{\mbox{\tiny TM}}$.
The hermitian structure reads: $\big<\cdot,\cdot\big>_{\!\mbox{\tiny$\mathcal{E}_1$}} :=
\big<\cdot,\cdot\big>_{\!\mbox{\tiny S}}\,g_{\mbox{\tiny M}}$.

The {\it Clifford extension} of \eqref{twistedspinbdl} is denoted by (c.f. \cite{Tolksdorf1})
\beq
\label{twistedspinbdl2}
\pi_{\mbox{\tiny$\mathcal{E}$}}:\,\mathcal{E} := 
\mathcal{E}_1\ot_{\mbox{\tiny M}}Cl_{\mbox{\tiny M}}\longrightarrow M\,.
\eeq
Here, the grading involution and Clifford action, respectively, are given by 
$\tau_{\mbox{\tiny$\mathcal{E}$}} := \tau_{\mbox{\tiny$\mathcal{E}_1$}}\!\ot\tau_{\mbox{\tiny Cl}}$ 
and $\gamma_{\mbox{\tiny$\mathcal{E}$}} := 
\gamma_{\mbox{\tiny$\mathcal{E}_1$}}\!\ot\!{\rm Id}_{\mbox{\tiny Cl}}$.
The hermitian structure is $\big<\cdot,\cdot\big>_{\!\mbox{\tiny$\mathcal{E}$}} :=
\big<\cdot,\cdot\big>_{\!\mbox{\tiny$\mathcal{E}_1$}}\big<\cdot,\cdot\big>_{\!\mbox{\tiny Cl}}$. 

In what follows all vector bundles are regarded as complex vector bundles, though we do not
explicitly indicate complexifications.

We denote the covariant derivative of the spin connection by $\nabla^{\mbox{\tiny S}}$. The 
corresponding spin-Dirac operator is $\,{/\!\!\!\!\nabla}^{\mbox{\tiny S}}$. 


For $A\in\Omega^1(M,\Lambda^{\!2}T^\ast\!M)$, the most general metric connection on the tangent bundle is known to be given by the covariant derivative
\beq
\label{genmetriccon}
\nabla^{\mbox{\tiny$g$}} := \nabla^{\mbox{\tiny LC}} + A\,.
\eeq
Here, $\nabla^{\mbox{\tiny LC}}$ is the covariant derivative of the Levi-Civita connection on the tangent bundle with respect to $g_{\mbox{\tiny M}}$. 

Accordingly, the torsion $\tau_{\mbox{\tiny$\nabla^g$}}\in\Omega^2(M,TM)$ of a metric connection 
\eqref{genmetriccon} can be expressed by the {\it torsion form} (also called ``torsion tensor'')
\beq
\label{torbypot}
\tau_{\!\mbox{\tiny A}}(u,v) \equiv A(u)v - A(v)u\qquad(u,v\in TM)\,.
\eeq

Indeed, the definitions \eqref{genmetriccon} and \eqref{torbypot} imply that for all smooth tangent vector fields $X,Y\in{\goth S}ec(M,TM)$:
\beq
\begin{split}
\tau_{\!\mbox{\tiny A}}(X,Y) &= \nabla^{\mbox{\tiny$g$}}_{\!\!X}Y - \nabla^{\mbox{\tiny LC}}_{\!\!X}Y -
\nabla^{\mbox{\tiny$g$}}_{\!\!Y}X + \nabla^{\mbox{\tiny LC}}_{\!\!Y}X\\
&=
\nabla^{\mbox{\tiny$g$}}_{\!\!X}Y -
\nabla^{\mbox{\tiny$g$}}_{\!\!Y}X - [X,Y]\\
&=
d_ {\mbox{\tiny$\nabla^g$}}\goth{Id}(X,Y)\,.
\end{split}
\eeq
Hence, we do not make a distinction between the torsion 
$\tau_{\mbox{\tiny$\nabla^g$}}\in\Omega^2(M,TM)$ of a metric
connection $\nabla^{\mbox{\tiny$g$}}$ and the torsion form $\tau_{\!\mbox{\tiny A}}\in\Omega^2(M,TM)$ 
of $A\in\Omega^1(M,\Lambda^{\!2}T^\ast\!M)$. 

Notice that \eqref{torbypot} may also be inverted (c.f. Prop. 2.1 in \cite{Agricola} and the corresponding
references cited there; In particular, see also \cite{tricerrri et al}).\footnote{The author would like to thank Ch. Stephan (University of Potsdam/Germany) for making him aware of these references.} In fact, one 
has for all $u,v,w\in TM$:
\beq
2g_{\mbox{\tiny M}}(A(u)v,w) =
g_{\mbox{\tiny M}}(\tau_{\!\mbox{\tiny A}}(u)v,w) -
g_{\mbox{\tiny M}}(\tau_{\!\mbox{\tiny A}}(v)w,u) +
g_{\mbox{\tiny M}}(\tau_{\!\mbox{\tiny A}}(w)u,v)\,.
\eeq

\begin{definition}
Let $\nabla^{\mbox{\tiny$g$}}$ be the covariant derivative of a metric connection on the tangent bundle
of $(M,g_{\mbox{\tiny M}})$. We call the section 
\beq
\label{torpot}
A := \nabla^{\mbox{\tiny$g$}} - \nabla^{\mbox{\tiny LC}}\in\Omega^1(M,\Lambda^{\!2}T^\ast\!M)
\eeq 
the ``torsion potential'' of $\tau_{\!\mbox{\tiny A}} = d_ {\mbox{\tiny$\nabla^g$}}\goth{Id}$.
\end{definition}

Consider the following hermitian {\it Clifford connection} on the Clifford module bundle 
\eqref{twistedspinbdl} that is provided by the following covariant derivative:
\beq
\label{twistedspincon1}
\begin{split}
\da &:= \nabla^{\mbox{\tiny${\rm S}\!\ot\!{\rm TM}$}}
\equiv
\nabla^{\mbox{\tiny S}}\!\ot\!{\rm Id}_{\mbox{\tiny TM}} +
{\rm Id}_{\mbox{\tiny S}}\!\ot\!\nabla^{\mbox{\tiny$g$}}\\
&=
\nabla^{\mbox{\tiny S}}\!\ot\!{\rm Id}_{\mbox{\tiny TM}} +
{\rm Id}_{\mbox{\tiny S}}\!\ot\!\nabla^{\mbox{\tiny LC}} + {\rm Id}_{\mbox{\tiny S}}\!\ot\!A\\
&=:
\nabla^{\mbox{\tiny$\mathcal{E}_1$}} + {\rm Id}_{\mbox{\tiny S}}\!\ot\!A\\ 
&\equiv
\nabla^{\mbox{\tiny$\mathcal{E}_1$}} + A\,.
\end{split}
\eeq
Clearly, \eqref{twistedspincon1} is but the gauge covariant derivative of a twisted spin connection on 
\eqref{twistedspinbdl} which is defined by the lift of \eqref{genmetriccon}.

Of course, every metrical connection \eqref{genmetriccon} on $(M,g_{\mbox{\tiny M}})$ can be lifted 
to the spinor bundle $\pi_{\mbox{\tiny S}}:\,S\rightarrow M$. However, in this case the resulting spin connection is neither a Clifford connection in the sense of our definition \eqref{cliffcon1}, nor is the ordinary Dirac action real-valued. 

With respect to an oriented orthonormal frame $e_1,\ldots,e_n\in\goth{S}ec(U,TM)$, with the dual frame being denoted by $e^1,\ldots, e^n\in\goth{S}ec(U,T^\ast\!M)$, the corresponding {\it twisted spin-Dirac operator} reads:
\beq
\label{twistspindirop1}
\begin{split}
\dda &= \,{/\!\!\!\!\nabla}^{\mbox{\tiny S}}\ot{\rm Id}_{\mbox{\tiny TM}} + 
\sum_{k=1}^n\gamma_{\mbox{\tiny S}}(e^k)\ot\nabla^{\mbox{\tiny$g$}}_{\!\!e_k}\\
&=
{/\!\!\!\!\nabla}^{\mbox{\tiny S}}\ot{\rm Id}_{\mbox{\tiny TM}} + 
\sum_{k=1}^n\gamma_{\mbox{\tiny S}}(e^k)\ot
\nabla^{\mbox{\tiny LC}}_{\!\!e_k} + \sum_{k=1}^n\gamma_{\mbox{\tiny S}}(e^k)\ot A(e_k)\\
&\equiv
{/\!\!\!\!\nabla}^{\mbox{\tiny$\mathcal{E}_1$}} + \sum_{k=1}^n\gamma_{\mbox{\tiny S}}(e^k)\ot A(e_k) \equiv
{/\!\!\!\!\nabla}^{\mbox{\tiny$\mathcal{E}_1$}} + \,{/\!\!\!\!A}\,.
\end{split}
\eeq
It looks similar to the usual gauge covariant Dirac operator encountered in ordinary electrodynamics on Minkowski space-time.

Accordingly, on the Clifford extension \eqref{twistedspinbdl2} we consider the Clifford connection
\beq
\label{twistedspincon2}
\begin{split}
{\tilde\nabla}^{\mbox{\tiny$\mathcal{E}$}} &:= 
\da\!\ot{\rm Id}_{\mbox{\tiny Cl}} +
{\rm Id}_{\mbox{\tiny$\mathcal{E}_1$}}\!\ot\nabla^{\mbox{\tiny Cl}}\\
&\equiv
\nabla^{\mbox{\tiny$\mathcal{E}$}} + 
{\rm Id}_{\mbox{\tiny S}}\!\ot\!A\!\ot{\rm Id}_{\mbox{\tiny Cl}}\,,
\end{split}
\eeq
with $\nabla^{\mbox{\tiny Cl}}$ being the induced Levi-Civita connection on the Clifford bundle.

With regard to the canonical embedding 
$\pi_{\mbox{\tiny$\mathcal{E}$}}|_{\mbox{\tiny S}}:\,
\mathcal{E}_1\hookrightarrow\mathcal{E}\rightarrow M,\;z\mapsto z\equiv z\ot 1$ one obtains for 
$\psi\equiv\psi\ot 1\in{\goth Sec}(M,\mathcal{E})$ the equality
\beq
{\tilde\nabla}^{\mbox{\tiny$\mathcal{E}$}}\psi = \da\psi\ot 1\,.
\eeq

\begin{definition}
Let $d_{\nabla^{{\rm LC}}}$ be the exterior covariant derivative induced by the Levi-Civita connection
with the (Riemannian) curvature denoted by $F_{\mbox{\tiny$\nabla^{\rm LC}$}}$. Also, let 
$F_{\mbox{\tiny$\nabla^g$}}\in\Omega^2(M,\Lambda^{\!2}T^\ast\!M)$ be the curvature of
$\nabla^{\mbox{\tiny$g$}}$. We call the relative curvature
\beq
\label{torsion2form}
\begin{split}
F_{\!\mbox{\tiny A}} &:= F_{\mbox{\tiny$\nabla^g$}} - F_{\mbox{\tiny$\nabla^{\rm LC}$}} =
d_{\nabla^{{\rm LC}}}A + A\wedge A\\
&=
d_{\nabla^{{\rm LC}}}A + \mbox{\small$\frac{1}{2}$}[A,A]\in\Omega^2(M,\Lambda^{\!2}T^\ast\!M)
\end{split}
\eeq
the ``torsion field strength'' associated to the torsion potential 
$A = \nabla^{\mbox{\tiny$g$}} - \nabla^{\mbox{\tiny{\rm LC}}}$.
\end{definition}


Notice that 
on a metrical flat manifold $(M,g_{\mbox{\tiny M}})$ the torsion field strength fulfills a Bianchi identity and therefore becomes a true curvature that is defined by torsion. In some approaches to dynamical torsion (``teleparallel gravity'') this curvature is used to describe gravity on metrical flat space-time manifolds, sometimes called ``Weitzenb\"ock space-times''.

Let again $e_1,\dots,e_n\in\goth{S}ec(U,TM)$ be a local (oriented orthonormal) frame with the dual frame being denoted by $e^1,\dots,e^n\in\goth{S}ec(U,T^\ast\!M)$. Also, let
\beq
\label{ymchi}
\begin{split}
\Sigma &:= \sum_{b = 1}^n e^b\ot\Sigma_b\in\Omega^1(M,{\rm End}^-_\gamma(\mathcal{E}))\,,\\
\Sigma_b &:= 
\sum_{a = 1}^n{\rm Id}_{\mbox{\tiny S}}\!\ot F_{\!\mbox{\tiny A}}(e_b,e_a)\ot e^a\in
\mathcal{C}^\infty(U,{\rm End}^-_\gamma(\mathcal{E}))\,.
\end{split}
\eeq
Again, ${\rm End}^-_\gamma(\mathcal{E})\subset{\rm End}(\mathcal{E})$ denotes the sub-algebra of the (odd) endomorphisms which commute with the Clifford action provided by $\gamma_{\mathcal{E}}$.

We consider the Dirac operator of simple type
\beq
\label{tordirop}
\DDD := {\,/\!\!\!\!{\tilde\nabla}}^{\mbox{\tiny$\mathcal{E}'$}} + 
\tau_{\mbox{\tiny$\mathcal{E}'$}}\phi_{\mbox{\tiny D}}
\eeq
acting on sections of the {\it Clifford twist} (for this notion see \cite{Tolksdorf1}) 
\beq
\pi_{\mbox{\tiny$\mathcal{E}'$}}:\,\mathcal{E}' := \mathcal{E}\ot_{\mbox{\tiny M}}Cl_{\mbox{\tiny M}}\rightarrow M
\eeq 
of \eqref{twistedspinbdl2} (i.e. the {\it double twist} of \eqref{twistedspinbdl}). The grading involution, 
the Clifford action and the hermitian product are defined, respectively, by 
\beq
\tau_{\mbox{\tiny$\mathcal{E}'$}} := 
\tau_{\mbox{\tiny$\mathcal{E}$}}\!\ot\!{\rm Id}_{\mbox{\tiny Cl}}\,,\quad 
\gamma_{\mbox{\tiny$\mathcal{E}'$}} := 
\gamma_{\mbox{\tiny$\mathcal{E}$}}\!\ot\!{\rm Id}_{\mbox{\tiny Cl}}\,,\quad 
\big<\cdot,\cdot\big>_{\!\mbox{\tiny$\mathcal{E}'$}} :=
\big<\cdot,\cdot\big>_{\!\mbox{\tiny$\mathcal{E}$}}\big<\cdot,\cdot\big>_{\!\mbox{\tiny Cl}}\,.
\eeq

Also,
\beq
{\tilde\nabla}^{\mbox{\tiny$\mathcal{E}'$}} := 
{\tilde\nabla}^{\mbox{\tiny$\mathcal{E}$}}\!\ot{\rm Id}_{\mbox{\tiny Cl}} +
{\rm Id}_{\mbox{\tiny$\mathcal{E}$}}\!\ot\nabla^{\mbox{\tiny Cl}}
\eeq
is the covariant derivative of the induced Clifford connection on the Clifford twist of \eqref{twistedspinbdl2}.
Also, we put
\beq
\begin{split}
\phi_{\mbox{\tiny D}} &:= -\sum_{b = 1}^n\Sigma_b\ot e^b\\
&=
\sum_{a,b = 1}^n{\rm Id}_{\mbox{\tiny S}}\!\ot F_{\!\mbox{\tiny A}}(e_a,e_b)\ot e^a\ot e^b\in
{\goth Sec}(M,{\rm End}^-_\gamma(\mathcal{E}'))\,.
\end{split}
\eeq

Notice that the Dirac operator of simple type \eqref{tordirop} is fully determined by the most general
hermitian Clifford connection \eqref{twistedspincon1} on the twisted spinor bundle \eqref{twistedspinbdl}. 
In contrast, the quantization of the lift of $\nabla^{\mbox{\tiny$g$}}$ to the spinor bundle itself is neither 
a quantized Clifford connection, nor a Dirac operator of simple type.

\begin{proposition}\label{satz} 
Let $M$ be closed compact. When restricted to the class of simple type Dirac operators \eqref{tordirop} 
and to sections $\psi\in\goth{S}ec(M,\mathcal{E}_1)\subset\goth{S}ec(M,\mathcal{E}')$, the total Dirac action decomposes as
\beq
\label{totdiraction}
\mathcal{I}_{\mbox{\tiny D,tot}}(\,\DDD,\psi) =
\int_M\!\ast\Big(-\epsilon\mbox{\small$\frac{rk(\mathcal{E}')}{4}$}scal(g_{\mbox{\tiny M}}) +
\big<\psi,\,\dda\psi\big>_{\!\mbox{\tiny$\mathcal{E}_1$}} - 
2^{2n}rk(S)\|F_{\!\!\mbox{\tiny A}}\|^2\Big)\,,
\eeq
where 
$\|F_{\!\!\mbox{\tiny A}}\|^2 \equiv -g^\ast_{\mbox{\tiny M}}(e^a,e^c)\,g^\ast_{\mbox{\tiny M}}(e^b,e^d)\,
tr\big(F_{\!\!\mbox{\tiny A}}(e_a,e_b)F_{\!\mbox{\tiny A}}(e_c,e_d)\big) \equiv
-trF_{\!ab}F^{ab}\in\mathcal{C}^\infty(M)$\,.

The variation of \eqref{totdiraction} with respect to the torsion potential 
$A\in\Omega^1(M,\Lambda^{\!2}T^\ast\!M)$ yields a Yang-Mills like equation for the torsion field strength:
\beq
\label{ymtoreq}
\delta_{\nabla^{\!g}}F_{\!\!\mbox{\tiny A}} = -\lambda_0 Re\big<\psi,\Theta\psi\big>_{\!\mbox{\tiny S}}\,.
\eeq
Here, $\lambda_0 \equiv 2^{-2(n+1)} \varepsilon n/rk(S)$ is a constant and
$\delta_{\nabla^g}\equiv (-1)^{n(k+1) + q + 1}\ast\! d_{\nabla^{\mbox{\tiny$g$}}}\ast$ is the formal
adjoint of the exterior covariant derivative induced by $\nabla^{\mbox{\tiny$g$}}$. 

When an oriented orthonormal frame is used, the right-hand side of \eqref{ymtoreq} explicitly reads: 
\beq
\label{spincurrent}
\begin{split}
Re\big<\psi,\Theta\psi\big>_{\!\mbox{\tiny S}} &:= 
\frac{\epsilon}{n}\sum_{i,j,k,l,m=1}^ng_{\mbox{\tiny M}}(e_i,e_j)g_{\mbox{\tiny M}}(e_l,e_k)
Re\big<\psi^k,\gamma_{\mbox{\tiny S}}(e^j)\psi^m\big>_{\!\!\mbox{\tiny S}}\,e^i\ot e^l\ot e_m\\
&=
\frac{\epsilon}{n}\sum_{i,j,k=1}^n Re\big<\psi_i,\gamma_k\psi^j\big>_{\!\!\mbox{\tiny S}}\,e^k\ot e^i\ot e_j\\
&=
\frac{\epsilon}{n}\sum_{i,j,k=1}^n\big<\psi_i,\gamma_k\psi_j\big>_{\!\!\mbox{\tiny S}}\,e^k\ot e^i\wedge e^j\\
&\equiv
\frac{\epsilon}{n}\,\goth{J}_{\mbox{\tiny spin}}\in\Omega^1(M,\Lambda^{\!2}T^\ast\!M)\,,
\end{split}
\eeq
whereby $\psi =: \sum_{k=1}^n\psi^k\!\ot\!e_k\in{\goth S}ec(M,\mathcal{E}_1)$ and
$\psi_i \equiv \sum_{j=1}^n g_{\!\mbox{\tiny M}}(e_i,e_j)\psi^j\in{\goth S}ec(U,S)$ are locally defined
spinor fields for all $i = 1,\ldots, n$.
\end{proposition}

Before we prove the Proposition \eqref{satz} it might be worthwhile adding some general comments first: 
When the embedding 
${\rm End(TM)} \simeq{\rm End}_\gamma(\mathcal{E}_1)\hookrightarrow {\rm End}(\mathcal{E}_1),\;\goth{B}\mapsto{\rm id}_{\mbox{\tiny S}}\!\ot\!{\goth B}$ is taken into account, one may replace \eqref{ymtoreq} by
\beq
\label{ymtoreq2}
\delta_{\!\mbox{\tiny A}}F_{\!\mbox{\tiny A}} = -\lambda_0 Re\big<\psi,\Theta\psi\big>_{\!\mbox{\tiny S}}\,,
\eeq
with $\delta_{\!\mbox{\tiny A}}$ being the formal adjoint of the exterior covariant derivative 
$d_ {\!\mbox{\tiny A}}$ of the defined by connection \eqref{twistedspincon1}. In the form \eqref{ymtoreq2}, the field equation \eqref{ymtoreq} for the torsion looks even more similar to the ordinary (non-abelian) Yang-Mills equation, although $d_{\!\mbox{\tiny A}}F_{\!\mbox{\tiny A}} \not=0$ unless 
$(M,g_{\mbox{\tiny M}})$ is flat.

Clearly, the functional \eqref{totdiraction} looks much like the usual Dirac-Yang-Mills action including
gravity. 
However, the energy-momentum current not only depends on the metric but also on its first derivative. 
This additional dependence, however, can be always (point-wise) eliminated by the choice of 
{\it normal coordinates} (nc.), such that at $x\in M$: 
\beq
F_{\!\mbox{\tiny A}}|_x \stackrel{\mbox{\tiny nc.}}{=} \big(dA + \mbox{\small$\frac{1}{2}$}[A,A]\big)|_x\,.
\eeq
Such a choice of local trivialization of the frame bundle does not affect the torsion potential 
(in contrast to the action of diffeomorphisms on $M$).

In order to end up with the field equation \eqref{ymtoreq} the sections 
$\psi\in{\goth S}ec(M,\mathcal{E}_1)$ are twisted fermions of spin 3/2, as opposed to ordinary 
Dirac-Yang-Mills theory. In the presented approach to torsion the additional spin-one degrees of 
freedom of matter are regarded as ``internal gauge degrees'' that couple to torsion.
We stress that the functional \eqref{totdiraction} is real-valued, indeed. This is because, the Dirac operator
$\dda$ is symmetric and the twisted spin-connection provided by $\nabla^{\mbox{\tiny$\mathcal{E}_1$}}$ 
is torsion-free. If the spinor bundle were not twisted with the tangent bundle in 
\eqref{twistedspinbdl}, then one has to use \eqref{genmetriccon} instead of 
$\nabla^{\mbox{\tiny$\mathcal{E}_1$}}$ to define the ordinary Dirac action. In this case, however, the functional \eqref{totdiraction} would be complex, in general, as mentioned already. The demand to derive a {\it real action including (full) torsion} from Dirac operators of simple type basically dictates the geometrical setup presented that eventually leads to \eqref{totdiraction}.

\Proof 
The statement of \eqref{satz} is a special case of the Proposition 6.2 in \cite{Tolksdorf1}.
To prove the statement we consider the $\zz_2-$graded hermitian vector bundle 
$\pi_{\mbox{\tiny E}}:\, E := 
TM\!\ot_{\mbox{\tiny M}} Cl_{\mbox{\tiny M}}\ot_{\mbox{\tiny M}} Cl_{\mbox{\tiny M}}\rightarrow M$. The grading involution and hermitian structure are given, respectively,  by
$\tau_{\mbox{\tiny E}} := {\rm Id}_{\mbox{\tiny TM}}\ot\tau_{\mbox{\tiny Cl}}\ot{\rm Id}_{\mbox{\tiny Cl}}$
and $\big<\cdot,\cdot\big>_{\!\mbox{\tiny E}} := g_{\mbox{\tiny M}}\big<\cdot,\cdot\big>_{\!\mbox{\tiny Cl}}\big<\cdot,\cdot\big>_{\!\mbox{\tiny Cl}}$. Hence, 
$\pi_{\mbox{\tiny$\mathcal{E}'$}}:\,\mathcal{E}' = S\ot_{\mbox{\tiny M}}E\rightarrow M$ is an odd twisted hermitian spinor bundle, with the grading involution $\tau_{\mbox{\tiny$\mathcal{E}'$}} :=
\tau_{\mbox{\tiny S}}\ot\tau_{\mbox{\tiny E}}$ and the Clifford action provided by 
$\gamma_{\mbox{\tiny$\mathcal{E}'$}} := \gamma_{\mbox{\tiny S}}\ot{\rm Id}_{\mbox{\tiny E}}$. The 
hermitian structure is $\big<\cdot,\cdot\big>_{\!\mbox{\tiny$\mathcal{E}'$}} := 
\big<\cdot,\cdot\big>_{\!\mbox{\tiny S}}\big<\cdot,\cdot\big>_{\!\mbox{\tiny E}}$. Furthermore, the twisted
spinor bundle carries the canonical Clifford connection that is provided by 
$\nabla^{\mbox{\tiny$\mathcal{E}'$}} = \nabla^{\mbox{\tiny$S\!\ot\!E$}}$, where
$\nabla^{\mbox{\tiny E}} := \nabla^{\mbox{\tiny${\rm TM}\!\ot\!{\rm Cl}\!\ot\!{\rm Cl}$}}$ and
$\nabla^{\mbox{\tiny TM}}\equiv\nabla^{\mbox{\tiny LC}}$.

From the general statement concerning the universal Dirac action restricted to Dirac operators of simple type it follows that
\beq
\mathcal{I}_{\mbox{\tiny D}}(\,\DDD) = 
\int_M\ast tr_{\!\gamma}\big(curv({\,/\!\!\!\!\nabla}^{\mbox{\tiny$\mathcal{E}'$}}) + 
tr_{\!\mbox{\tiny$\mathcal{E}'$}}\phi_{\mbox{\tiny D}}^2\big)\,.
\eeq

One infers from the ordinary Lichnerowicz-Schr\"odinger formula of twisted spin-Dirac operators 
(c.f. \cite{Lichnerowicz} and \cite{Schroedinger}) that
\beq
tr_{\!\gamma}curv({\,/\!\!\!\!\nabla}^{\mbox{\tiny$\mathcal{E}'$}}) = 
-\epsilon\mbox{\small$\frac{rk(\mathcal{E}')}{4}$}scal(g_{\mbox{\tiny M}})\,.
\eeq

Furthermore, it is straightforward to check that
\beq
tr_{\!\mbox{\tiny$\mathcal{E}'$}}\phi_{\mbox{\tiny D}}^2 \thicksim\|F_{\!\mbox{\tiny A}}\|^2\,.
\eeq

Finally, when restricting to sections 
$\psi\in\goth{S}ec(M,\mathcal{E}_1)\subset\goth{S}ec(M,\mathcal{E}')$ one obtains
\beq
\label{gaugecovdirlag}
\begin{split}
\big<\psi,\,\DDD\psi\big>_{\!\mbox{\tiny$\mathcal{E}'$}} &= 
\big<\psi,\,\dda\psi\big>_{\!\mbox{\tiny$\mathcal{E}_1$}}\\
&=
\big<\psi,\,{/\!\!\!\!\nabla}^{\mbox{\tiny$\mathcal{E}_1$}}\psi\big>_{\!\mbox{\tiny$\mathcal{E}_1$}} + 
\big<\psi,\,{/\!\!\!\!A}\psi\big>_{\!\mbox{\tiny$\mathcal{E}_1$}}\,.
\end{split}
\eeq
The first equality holds because $\bigoplus_{k\geq 0}\Lambda^{\!k}T^\ast\!M$ is an orthogonal sum.
Hence,
\beq
\label{fermlagrange}
\begin{split}
\big<\psi,\tau_{\mbox{\tiny$\mathcal{E}'$}}\phi_{\mbox{\tiny D}}\psi\big>_{\!\mbox{\tiny$\mathcal{E}'$}} &=
\sum_{a,b=1}^n\big<\psi,(\tau_{\mbox{\tiny S}}\ot F_{\!\mbox{\tiny A}}(e_a,e_b))\psi\big>_{\!\mbox{\tiny$\mathcal{E}_1$}}\big<1,e^a\big>_{\!\mbox{\tiny Cl}}\big<1,e^b\big>_{\!\mbox{\tiny Cl}}\\
&=
0\,.
\end{split}
\eeq
This proves \eqref{totdiraction}. To also prove \eqref{ymtoreq} we remark that up to a boundary term
\beq
\|F_{\!\mbox{\tiny A}}\|^2 \thicksim \big<A,\delta_{\nabla^{\mbox{\tiny$g$}}}F_{\!\mbox{\tiny A}}\big>_{\!\mbox{\tiny$\Lambda{\rm T}^\ast\!{\rm M}\!\ot\!\Lambda^{\!2}{\rm T}^\ast\!{\rm M}$}}\,.
\eeq
This is similar to the usual Yang-Mills-Lagrangian. The basic difference is that the torsion potential
$A\in\Omega^1(M,\Lambda^{\!2}T^\ast\!M)$ itself does not define a connection, in general. As already mentioned, if $(M,g_{\mbox{\tiny M}})$ is flat, then the torsion field strength $F_{\!\mbox{\tiny A}}$ has the geometrical meaning of the curvature of a general metric connection parametrized by torsion potentials. 
In any case, together with the solutions of the coupled Einstein-Dirac equation, the solutions of 
\eqref{ymtoreq} determine the torsion of a general metric connection. \QED

We remark that by an appropriate re-definition of the sections $\phi$ and $\psi$ one may always recast 
the functional \eqref{totdiraction} into the even more suggestive form
\beq
\label{totdiractionres1}
\mathcal{I}_{\mbox{\tiny D,tot}}(\,\DDD,\psi) \thicksim
-\mbox{\small$\frac{\varepsilon}{\lambda_{\mbox{\tiny grav}}}$}\!\int_M\!\ast scal(g_{\mbox{\tiny M}}) +
\int_M\ast\big<\psi,\,\dda\psi\big>_{\!\mbox{\tiny$\mathcal{E}_1$}} - \mbox{\small$\frac{1}{2{\rm g}^2}$}\!
\int_M tr\big(F_{\!\mbox{\tiny A}}\wedge\ast F_{\!\mbox{\tiny A}}\big)\,,
\eeq
with ${\rm g} > 0$ being an arbitrary (coupling) constant like in ordinary non-abelian Yang-Mills theory. Furthermore, by re-scaling the torsion potential $A$, which is admissible since the torsion potential belongs to a vector space as opposed to true gauge potentials, the field equation \eqref{ymtoreq2} changes to
\beq
\label{toreq}
\begin{split}
\delta_{\!\mbox{\tiny A}}F_{\!\mbox{\tiny A}} &= -{\rm g}\,\goth{J}_{\mbox{\tiny spin}}\,,
\end{split}
\eeq
with the spin-current $\goth{J}_{\mbox{\tiny spin}}\in\Omega^1(M,\Lambda^{\!2}T^\ast\!M)$ being
defined by \eqref{spincurrent}. The Dirac equation becomes
\beq
\label{direq}
\dda\psi = 0\quad\Leftrightarrow\quad 
{/\!\!\!\!\nabla}^{\mbox{\tiny$\mathcal{E}_1$}}\psi = -{\rm g}\,{/\!\!\!\!A}\psi\,.
\eeq


As in ordinary general relativity the metric (and thus the connection 
$\nabla^{\mbox{\tiny$\mathcal{E}_1$}}$) is determined by the Einstein equation with 
the energy-momentum current similarly defined as in the usual Dirac-Yang-Mills theory. 
The Einstein equation together with the Dirac-Yang-Mills like equation thus provide a closed system
to dynamically describe all classically admissible degrees of freedom introduced by the derived
action \eqref{totdiractionres1}. In particular, the solutions of these equations determine the Dirac
operator $\,\DDD$, as an extremum of the universal Dirac action, when the latter is restricted to the 
class of simple type Dirac operators \eqref{tordirop}. Notice that in the case considered the Dirac 
operators \eqref{tordirop} are fully determined by the metric and the torsion, that is to say by a general
metric connection $\nabla^{\mbox{\tiny$g$}}$ on the tangent bundle.

\section{Concluding remarks and outlook}
On the twisted spinor bundle \eqref{twistedspinbdl} there exists no Dirac operator of simply type other
than the quantized spin connection. The (double) twist of \eqref{twistedspinbdl} not only guarantees the existence of more general Dirac operators of simple type but also that the (relative) curvature added to the quantized connection \eqref{twistedspincon1} does not change the Dirac equation. In fact, it is well-known that the coupling of the fermions to curvature like terms may spoil renormalizability. Though the action 
\eqref{totdiractionres1} is derived by the distinguished class of Dirac operators \eqref{tordirop}, the inner degrees of freedom of matter only ``minimally couple'' to torsion similar to ordinary Dirac-Yang-Mills gauge
theory. Accordingly, the Yang-Mills like field equation \eqref{toreq} for torsion should be contrasted with 
the field equation \eqref{torpalatini} of ordinary Einstein-Cartan theory. The coupling constant 
${\rm g} > 0$ determines the coupling strength of the {\it inner degrees} of freedom of the fermions to torsion similar to ordinary Dirac-Yang-Mills theory. Like in Yang-Mills theory, the coupling 
constant is {\it dimensionless} in four (space-time) dimensions. In particular, it is independent 
of the gravitational constant. This is in strong contrast to the coupling constant obtained by the Palatini formalism of general relativity. Therefore, the assumption $0 < {\rm g} << 1$ makes the geometrical
description \eqref{twistedspinbdl} of matter in terms of Rarita-Schwinger fields phenomenological 
acceptable. Technically, such a weak coupling also allows to treat the field equations perturbatively. In particular, one also obtains non-trivial solutions of \eqref{toreq}, even if the coupling to the fermions 
and the self-coupling of the torsion potential is neglected (e.g. whenever all other interactions are assumed to be much stronger than the interaction with the torsion field). This is also in strong contrast to \eqref{torpalatini}. In the weak coupling case one gets back ordinary Einstein-Dirac theory whereby matter propagates in a non-trivial background provided by solutions of the linearized wave equation 
$\Box_g A - d\delta_{\!g}A = 0$, with the (space-time) metric $g_{\mbox{\tiny M}}$ being determined as in ordinary Einstein-Dirac theory. Hence, the energy of the propagating torsion $\tau_{\!\mbox{\tiny A}}$ is seen to contribute effectively to the cosmological constant and thus to the dark energy in the universe.
We stress that all of these considerations make physically sense even if gravitational effects are supposed to be negligible, such that $(M,g_{\mbox{\tiny M}})$ is Minkowski space-time.

In the scheme presented torsion couples to matter similar to Yang-Mills gauge theory. Hence, when Minkowski space-time is presumed one may apply an analogous procedure to quantize the torsion potential (and thus torsion) as for Yang-Mills gauge potentials. However, in four dimensions and signature $s = \mp 2$, the ``gauge group'' is provided by the Lorentz group and hence is non-compact. This is known to cause serious trouble, for instance, when one tries to make use of the path integral procedure widely used to (perturbatively) quantize non-abelian gauge theories. 

In a forthcoming work we shall discuss the influence of the torsion within the physical frame provided by the Standard Model when the latter is geometrically described by Dirac operators of simple type similar to the scheme presented here.  Especially, we shall discuss a possible influence on the Higgs mass when torsion is assumed to be massive. In \cite{thum et al} it has been shown how the Standard Model allows to make a prediction of the Higgs mass if the Standard Model is geometrically described by Dirac operators of simple type. Though the derived Higgs mass is within the range allowed by the Standard Model it is yet too big
than experimentally confirmed by the LHC. The coupling to torsion, however, introduces a possible additional degree of freedom and therefore may allow to lower the predicted value of the Higgs mass
in  \cite{thum et al} 
within the geometrical frame presented here, which slightly differs from that in loc. site.

\bibliographystyle{amsplain}
\def\dbar{\leavevmode\hbox to 0pt{\hskip.2ex \accent"16\hss}d}
\providecommand{\bysame}{\leavevmode\hbox to3em{\hrulefill}\thinspace}
\providecommand{\MR}{\relax\ifhmode\unskip\space\fi MR }

\providecommand{\MRhref}[2]{%
  \href{http://www.ams.org/mathscinet-getitem?mr=#1}{#2}
}
\providecommand{\href}[2]{#2}

\end{document}